\def\year{2019}\relax
\documentclass[letterpaper]{article}

\usepackage{aaai19}
\usepackage{times}
\usepackage{helvet}
\usepackage{courier}
\usepackage[table]{xcolor}
\usepackage[utf8]{inputenc}

\usepackage{subcaption}
\usepackage{mathtools}
\usepackage{amsmath}
\usepackage{amsthm}
\usepackage{amssymb}
\usepackage{multicol}
\usepackage{paralist}
\usepackage{amsfonts}
\usepackage{bbm}
\usepackage{graphics}      
\usepackage{txfonts}
\usepackage{color}
\usepackage{booktabs}
\usepackage{textcomp}
\usepackage{graphicx}
\usepackage{caption}
\usepackage{subcaption}
\usepackage{balance}  
\usepackage{blindtext}
\usepackage{hyperref}
\usepackage{url}

\usepackage{xargs} 
\usepackage[colorinlistoftodos,prependcaption,textsize=tiny]{todonotes}
\newcommandx{\ak}[2][1=]{\todo[linecolor=red,backgroundcolor=red!25,bordercolor=red,#1]{#2}}
\newcommandx{\vicenc}[2][1=]{\todo[linecolor=blue,backgroundcolor=blue!25,bordercolor=blue,#1]{#2}}
 \newcommandx{\cpc}[2][1=]{\todo[linecolor=blue,backgroundcolor=orange!25,bordercolor=orange,#1]{#2}}

\begin{document}

\title{Sharing emotions at scale: The Vent dataset}

\author{Nikolaos Lykousas\textsuperscript{1}~Constantinos Patsakis\textsuperscript{1}~Andreas Kaltenbrunner\textsuperscript{2}~Vicen\c{c} G\'omez\textsuperscript{2}\\\\
\textsuperscript{1}University of Piraeus, Greece~~~~\textsuperscript{2}Universitat Pompeu Fabra. Barcelona, Spain}

\maketitle
\begin{abstract}
The continuous and increasing use of social media has 
enabled the expression of human thoughts, opinions, and everyday actions publicly at an unprecedented scale. 
We present the Vent dataset, 
the largest annotated dataset of text, emotions, and social connections to date.
It comprises more than 33 millions of posts by nearly a million users together with their social connections. Each post has an associated emotion. There are 705 different emotions, organized in 63 ``emotion categories'', forming a two-level taxonomy of affects.
Our initial statistical analysis describes the global patterns of activity in the Vent platform, revealing large heterogeneities and certainly remarkable regularities regarding the use of the different emotions.
We focus on the aggregated use of emotions, the temporal activity, and the social network of users, and outline possible methods to infer emotion networks based on the user activity.
We also analyze the text and describe the \emph{affective landscape} of Vent, finding agreements with existing (small scale) annotated corpus in terms of emotion categories and positive/negative valences.
Finally, we discuss possible research questions that can be addressed from this unique dataset.

\end{abstract}

\section{Introduction}



Experiencing emotions is an integral part of human life, which plays significant role in the effective communication of people \cite{barrett2006solving}. In fact, sometimes, emotional intelligence is considered more important than cognitive intelligence for successful interaction \cite{pantic2005affective}. 
Naturally, humans have an innate need to share the emotions and feelings they experience, a phenomenon known as \emph{``social sharing of emotions''} \cite{rime1991beyond}. In the era of social networking, people extensively share what they feel via social media content to express their emotional experiences, reduce dissonance, deepen social connections, and convey their evaluations on a given topic \cite{berger2012makes}. Given their importance, a lot of research in the field of affective computing has focused on efficiently recognising emotions in online user behavior \cite{politou2017survey}.
To facilitate research in this field, we collected the data shared in Vent, a social network where users can \emph{``express and share their feelings with people who care''}\footnote{\url{https://www.vent.co/}}. In this work, we discuss several features and insights of this \textbf{complete} dataset \cite{lykousas_nikolaos_2019_2537838}.


Vent is a semi-anonymous social networking app that lets users share their feelings and frustrations without the fear of a negative backlash. It encourages users (also referred to as \textit{venters}) to voice their opinion to a supportive community without the worry of being insulted, de-friended or upsetting people they know. Although users must register to use the app, verification is not required, and there is no option for users to connect other social accounts, thus leaving room for anonymity. 

\begin{figure}

\centering
  \begin{subfigure}[t]{0.49\columnwidth}
  	\centering
    \includegraphics[width=\textwidth]{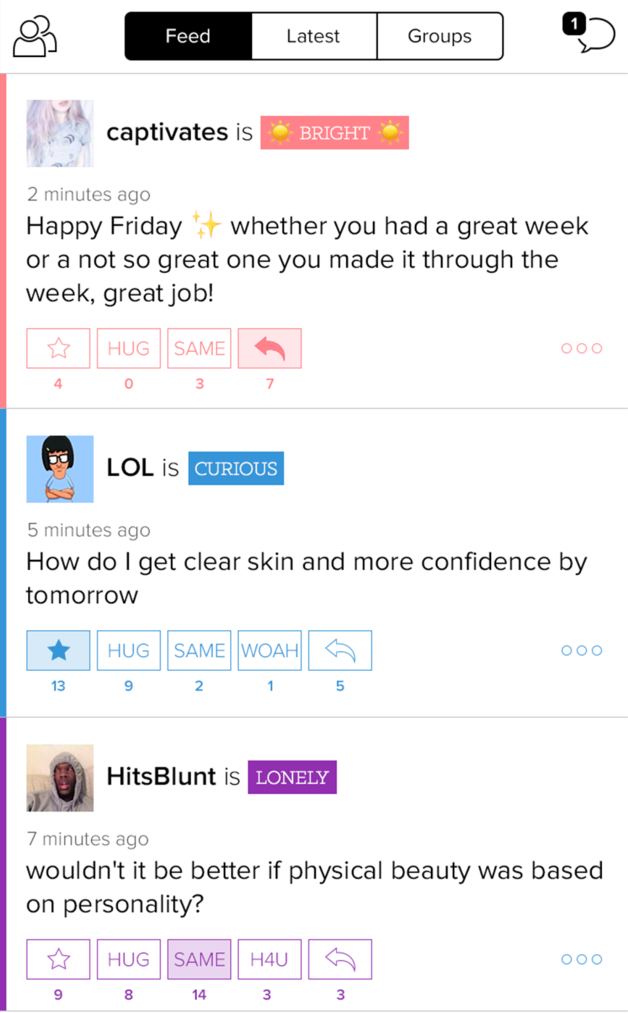}
    \caption{Vent Feed}
    \label{fig:vent_feed}
  \end{subfigure}
  \begin{subfigure}[t]{0.49\columnwidth}
  	\centering
    \includegraphics[width=\textwidth]{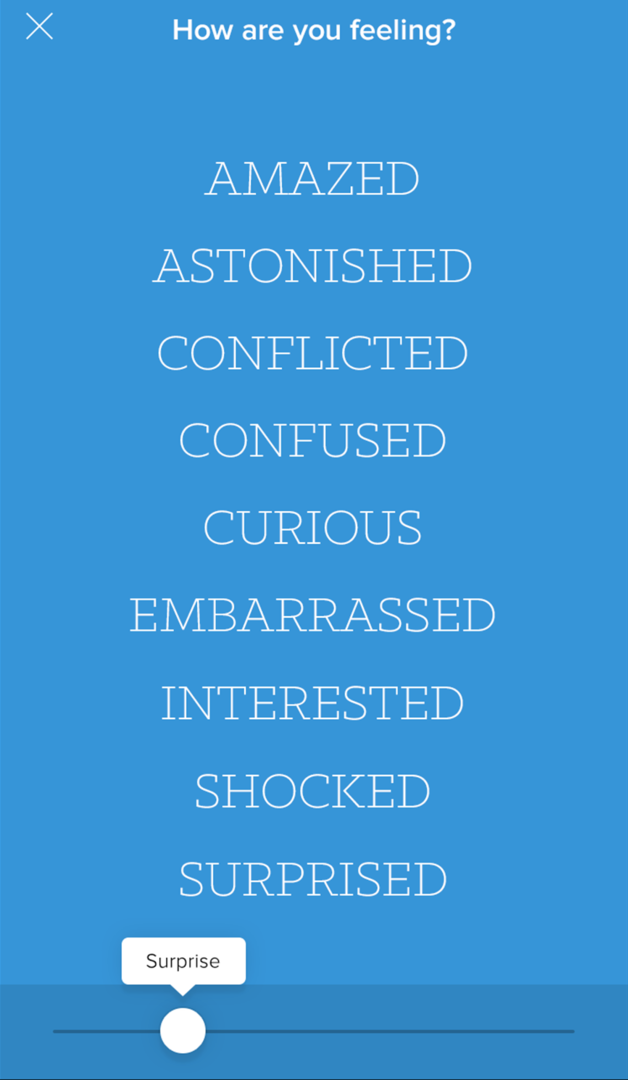}
    \caption{Emotion Picker}
    \label{fig:vent_emo_picker}
  \end{subfigure}
  \caption{Screenshots of the interface of Vent app.}
  \label{fig:vent_screen}
\end{figure}

Each post (also referred to as \textit{vent}) in the app is associated with a specific emotion by the user who submits it (Figure~\ref{fig:vent_emo_picker}).  ``\textit{Emotions}'' in the Vent platform are a fuzzy concept, probably better described as ``affects'', since they include a broad range of feelings, emotions and moods. Apart from posting, users can browse through the feeds of other users' vents (see Figure~\ref{fig:vent_feed}), and interact with them by commenting or reacting to their vents via a set of emotion-specific reactions (e.g. hug, same, h4u - here for you). Moreover, users can follow others and get updates on their vents, create and join groups, and exchange direct messages with the users they follow.

Maybe the most important feature of our dataset is the existence of self-reported ``ground truth'' affect annotations associated with each text. The unstructured and noisy nature of user-generated content on the Internet \cite{6468032}, along with the scarcity of ground truth labels regarding affects, consist a major challenge in the field of affective computing, particularly for tasks such as emotion detection and analysis. As highlighted in
Mohammad et al.~\citeyear{mohammad2018semeval}:
\begin{quote}
    \textit{It is challenging to obtain consistent annotations for affect due to a number of reasons, including: the subtle ways in which people can express affect, fuzzy boundaries of affect categories, and differences in human experience that impact how they perceive emotion in text.}
\end{quote}

In the realm of social media, the closely related problem of sentiment analysis has received significant attention, with the bulk of the literature focusing around social networks such as Twitter \cite{pak2010twitter,kouloumpis2011twitter,agarwal2011sentiment} and Facebook \cite{ortigosa2014sentiment,wang2012system,ahkter2010sentiment,troussas2013sentiment}, mainly due to their market share and wide availability of data.

However emotions are much more expressive than sentiments, and the current approaches for emotion analysis have a long way to go before matching the success and ubiquity of sentiment analysis \cite{wang2015detecting}. Nonetheless, the amount of useful information which can be gained by moving past the negative and positive sentiments and towards identifying discrete emotions can help improve many applications.

To this end, a variety of affective lexica have been proposed to offer information about affect expressed in text at a fine level of granularity~\cite{liu2012survey}. Lexica-based approaches, however, have some limitations in the domain of emotion detection, e.g., the lack of coverage, especially in social media/micro-blogging context, and the inherent incapability of recognising sentences without keywords ~\cite{kao2009towards,mudinas2012combining}.

Other works take different approaches to analyze emotions online. For example, Garcia et al.~(\citeyear{garcia2016dynamics}) use principles of dynamical systems to study the emotional states of a small group of users during their participation in online discussions. 
Bazarova et al.~(\citeyear{bazarova2015social}) performed an experiment in which participants labelled the emotions of their interaction on Facebook. Xu et al.~(\citeyear{moods}) developed a chatbot for customer service, and their content analysis revealed that more than 40\% of the user requests were emotional.

Based on the above, we consider that the Vent dataset can provide a baseline corpus for emotion analysis of the user-generated text. To the best of our knowledge, this is the largest annotated dataset of texts with affects. Moreover, the labelling has been made by the authors of the texts who can classify their texts more accurately according to what they felt when they wrote it, as in \cite{bazarova2015social}. 
Table~\ref{tab:sample} shows an illustrative sample of vents and their associated emotions.
We argue that this may overcome many biases, give more insight on how feelings can be expressed in written speech.

\begin{table}
\centering
\caption{Some illustrative sample vents.}
\label{tab:sample}
    \rowcolors{2}{gray!35}{white}
	\begin{tabular}{p{.25\columnwidth}p{.65\columnwidth}}
    \toprule
    	{\bf Emotion} & {\bf Vent text} \\
        \midrule
        Sad & I hate fuckin every single person on this fuckin planet. Someone kill me pls \\
        Happy & Best day I have had in a long time :) \\
	    Frustrated & i wish it was as easy to forget someone as it is to get attached to them \\
	    Stressed & really can't deal with school again today \\
	    Anxious & constantly worried that my boyfriend will fall out of love with me \\
	    Supportive & Hey guys, just a lil note that it's okay to want attention, it's super natural and there's nothing wrong with it \\
	    Affectionate & boy I like called me princess He's so precious \\
        Disappointed & the Highschool life isn't really fun. \\
        Curious & Do really hairy people use shampoo on their body or shower gel? Does it need conditioner? \\
    \bottomrule
    \end{tabular}
\end{table}

\section{Data Collection}
\label{sec:data_collection}

The Vent platform is offered exclusively as a mobile application for iOS and Android.
To the best of our knowledge, there is no open-source client available at the time of writing, hence, we follow a similar method as in \cite{Siekkinen2016,lykousas2018adult} to analyze the network traffic between the app and the backend service. To this end, we employ an SSL-capable man-in-the-middle proxy between a mobile device with the Vent app installed and the Vent service, that acts as a transparent proxy.

The proxy intercepts the HTTPS requests sent by the mobile device and pretends to be the server to the client and the client to the server, enabling us to examine and log the requests and responses between the client app and the server. Based on them, we identified a set of APIs allowing us to collect data about emotions and emotion categories, vents, and user relationships. Some of the APIs had limitations regarding the amount of returned data. For example, the vent feed API returned results up to one month ago. To overcome this limitation, we focused our efforts on obtaining a complete list of usernames and gathering the vents of each user individually.

The collection of the usernames was made thanks to the searching mechanism of Vent. More precisely, when querying the username search API, we observed that the input query was matched from the starting character of each username. Moreover, the API provided unlimited pagination until no more usernames were beginning with the input query. Therefore, we provided a list of all the valid characters for usernames. More precisely, all characters of the English alphabet (both lower/upper case), and numbers 0-9. 

This procedure resulted in a set of $1,161,265$ distinct usernames, $50,217$ of which belonged to users with private profiles.
For each of the remaining $1,111,048$ public profiles, we collected the full set of posted vents and their directed social links (lists of followed/following users). Social links between private profile users were not accessible.

The obtained dataset is \emph{complete}, in a sense it  
 contains every vent posted and every social link (in any direction) of a user with a public profile since the genesis of Vent (oldest vent created at 17/12/2013) until 02/12/2018, the date at which we initiated the crawling procedure. Note that the crawling lasted approximately two weeks, meaning that the dataset also includes vents created in the meantime since the collection of data was carried out in batches.

\subsection{Ethical considerations}
The  used methodology, if efficiently implemented,  may collect a wide range of information, part of which could be sensitive. Nevertheless, by using the service, as stated in terms of service, one grants all users to view the shared content for their personal, non-commercial purposes. In this regard, all content for users of Vent is ``public''.

Despite the public nature of the shared data, we consider that data about individuals must be published only in an anonymized and/or aggregated form. Therefore, all direct identifiers of users have been removed, as well as shared URLs and usernames. In addition, all unique identifiers have been masked to prevent user linking. Finally, rather than publishing all the collected information, since this is a complete dataset, we have opted to 
remove the text from the public dataset to further guarantee user privacy as in \cite{Garimella2018WhatsAppDA}. Researchers who want to access the texts for research and non-commercial use are welcome to contact us.

\subsection{Structure of the dataset}

The crawled dataset contains $33,623,414$ vents in total, posted by $934,095$ users. These vents are annotated with a total of 705 emotions organized in 63 ``emotion categories'', concretely forming a two-level taxonomy of affects.

The provided dataset is structured in files each containing a different entity (i.e. emotion categories, emotions, vents and social links). Entities external to each file are cross-referenced via the anonymized  universally unique identifiers (UUIDs).

\begin{figure}[!t]
  \centering
  \includegraphics[width=0.9\columnwidth]{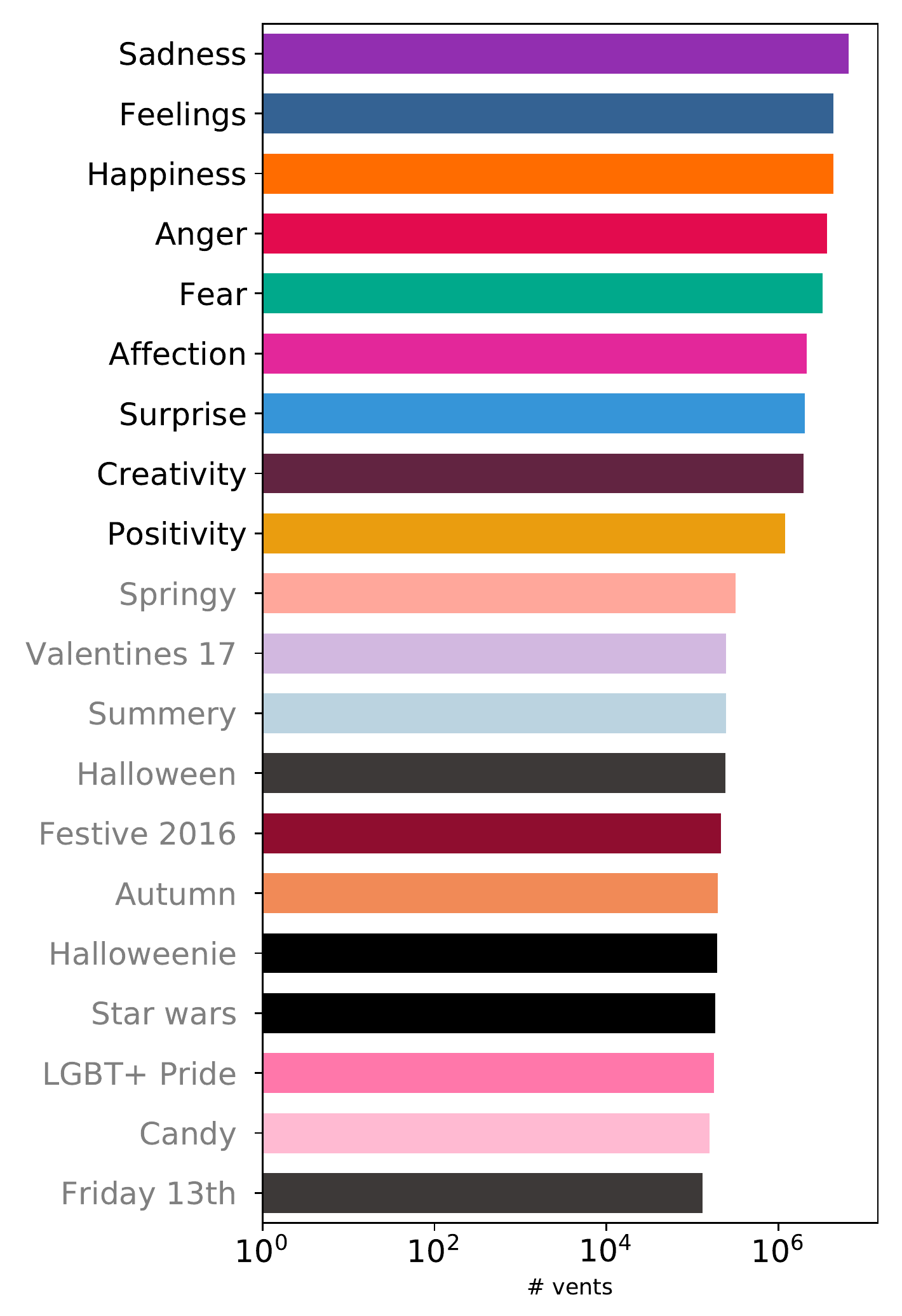}
  \caption{The twenty most used emotion categories (labels in grey indicates non-permanent categories).
	Colors for each category correspond to the ones provided in Vent app.}
  \label{fig:emo_cats}
\end{figure}

\begin{figure*}[!t]
  \centering
  \begin{subfigure}[t]{0.26\textwidth}
  	\centering
    \includegraphics[width=\textwidth]{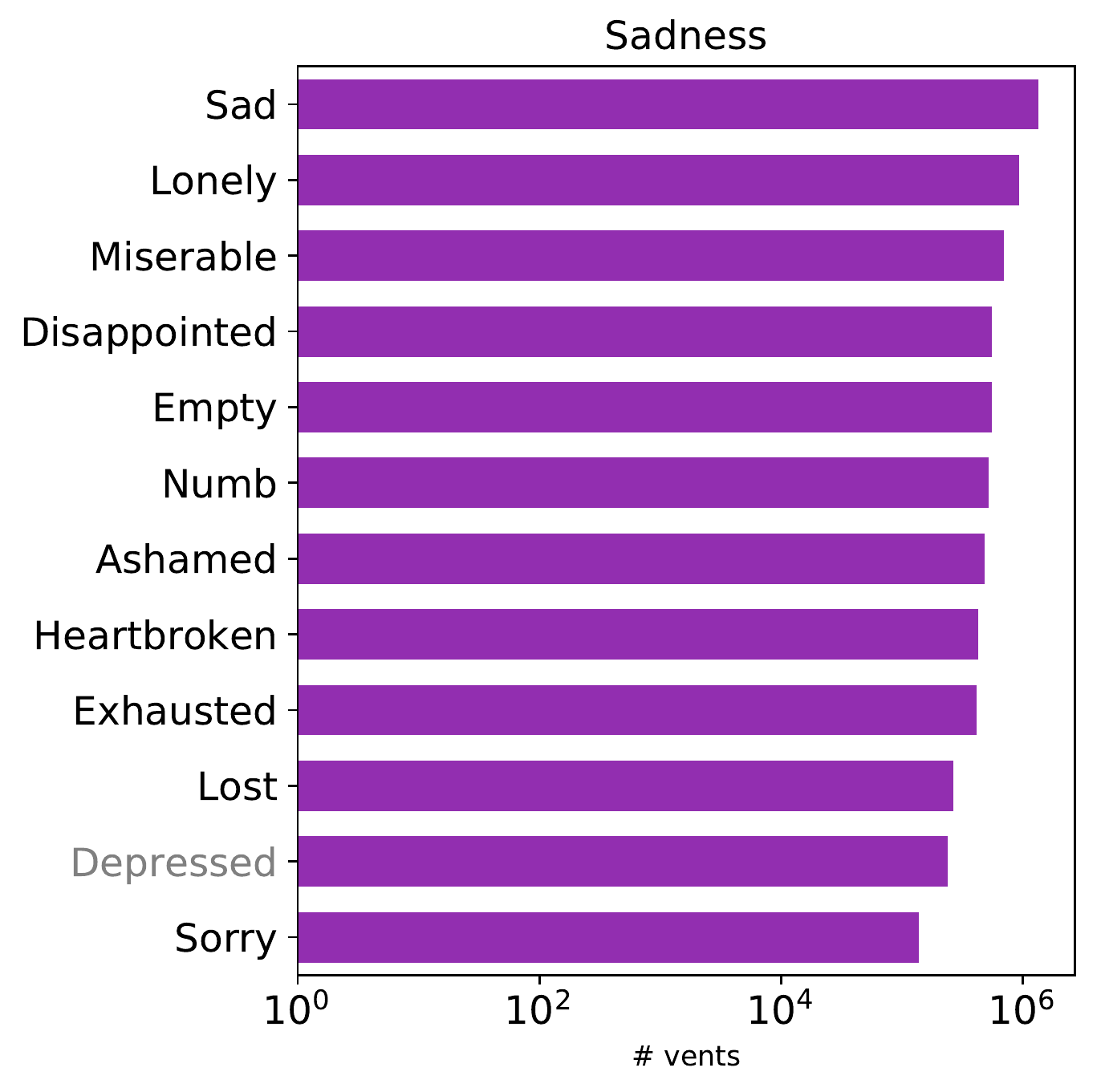}
    \caption{Category \emph{Sadness}}
  \end{subfigure}
  \begin{subfigure}[t]{0.28\textwidth}
  	\centering
    \includegraphics[width=\textwidth]{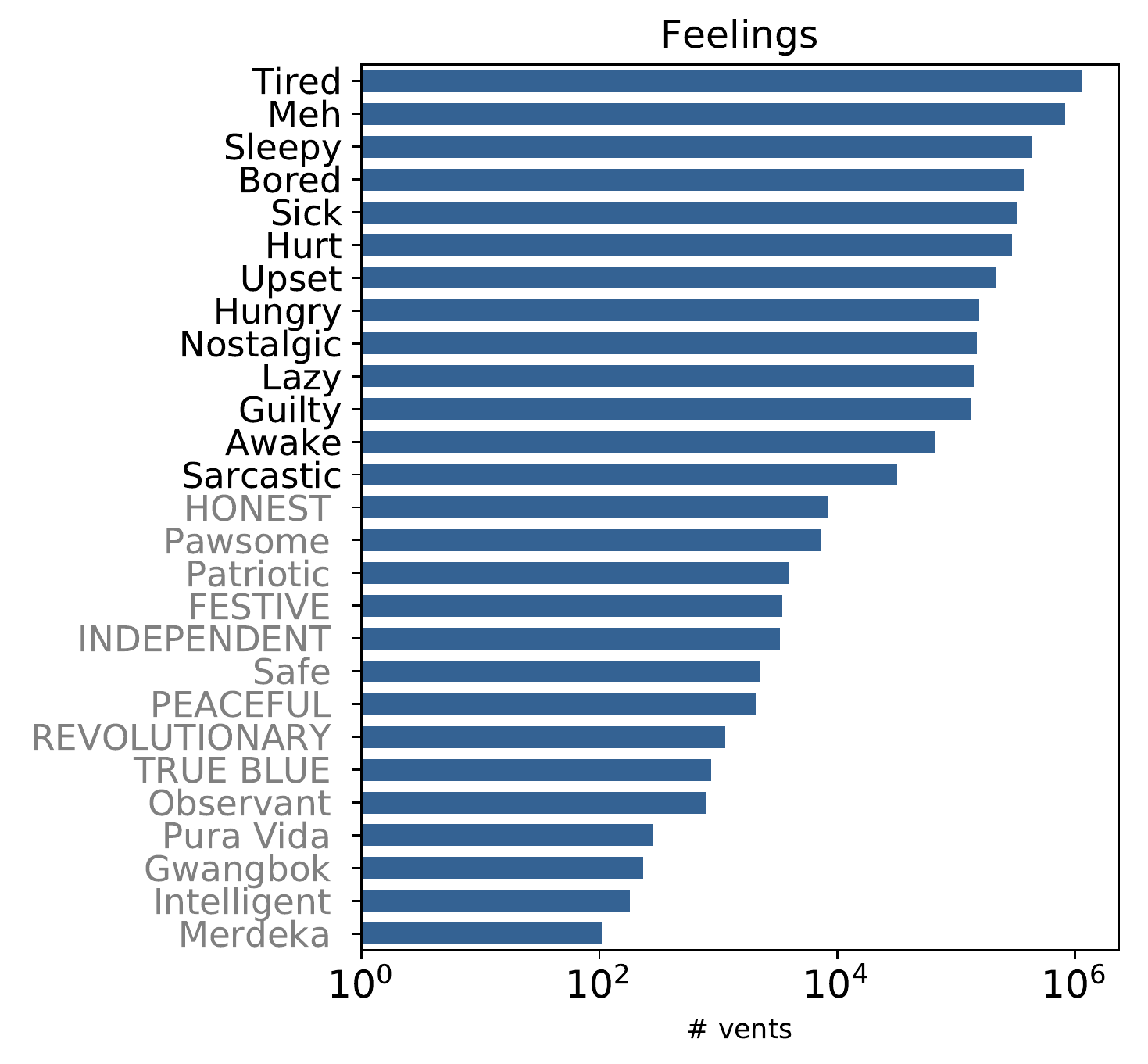}
    \caption{Category \emph{Feelings}}
  \end{subfigure}  
  \begin{subfigure}[t]{0.26\textwidth}
  	\centering
    \includegraphics[width=\textwidth]{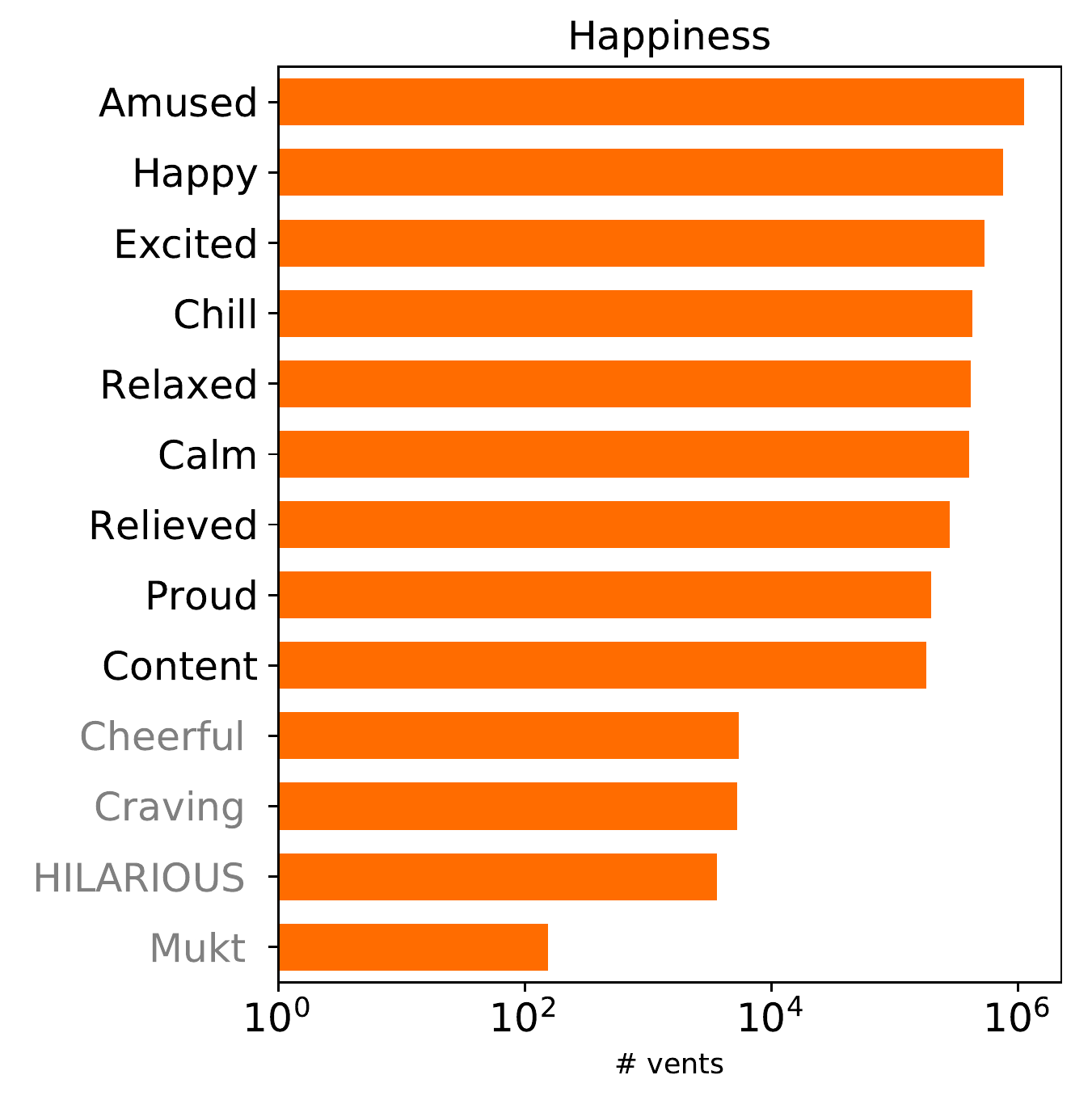}
    \caption{Category \emph{Happiness}}
  \end{subfigure}  
  \caption{Histograms of vents per emotion distributed per emotion categories. Disabled emotions are in gray.}
  \label{fig:emos}
\end{figure*}

Our full dataset consists of the following files and data:

\begin{compactitem}
\item  {\bf emotion\_categories.csv}:
\begin{compactitem}
\item \texttt{id} (string): a UUID associated with a specific emotion category.
\item \texttt{name} (string): The name of the emotion category.
\end{compactitem}

\item  \textbf{emotions.csv}:
\begin{compactitem}
\item \texttt{id} (string): a UUID associated with a specific emotion.
\item \texttt{emotion\_category\_id} (string): a UUID associated with the corresponding emotion category.
\item \texttt{name} (string): The name of the emotion.
\item \texttt{enabled} (boolean): Whether the emotion was enabled at the time of crawling. 
\end{compactitem}

\item  \textbf{vents.csv}:
\begin{compactitem}
\item \texttt{emotion\_id} (string): a UUID associated with a specific emotion (cross-reference to emotions.csv).
\item \texttt{user\_id} (string): a UUID associated with a specific user.
\item \texttt{created\_at} (string): Date when the vent was posted, in UTC. Provided as a string in the format of ``YYYY-MM-DD hh:mm:ss.sss''.
\item \texttt{reactions} (integer): Total number of reactions to a vent.
\item \texttt{text} (string): The raw textual content of each vent. To preserve the anonymity of our dataset and at the same time reduce noise, we replace user mentions and URLs found in vents, following an approach similar to the one described in \cite{ijca2018917319}.
We used a set of regular expressions to replace all URLs with the \_URL\_ token, and references to usernames with the \_USER\_REFERENCE\_ token. No other text processing tasks were performed. This file is available only upon request
as a restricted-access dataset
\footnote{\url{http://doi.org/10.5281/zenodo.2537982}}. Instead, in the publicly available dataset \cite{lykousas_nikolaos_2019_2537838}, we include a file named \textit{vents\_metadata.csv} which contains all the fields of vents.csv, except \textit{text}.
\end{compactitem}

\item  \textbf{vent.edges}: A snapshot of the social graph of Vent, at the time of crawling. It contains the directed friendship links between users.
The UUIDs of the users (nodes) are the same as in the file \textbf{vents.csv}.
\end{compactitem}

\section{Data Exploration}
\label{sec:data_exploration}
We now present an exploratory analysis of the Vent dataset.
We first look at the usage of the different emotions at the level of vent, emotions, and users.
We then briefly look at the temporal evolution of vents, the social network and outline a possible way to construct networks of emotions.
Finally, we analyze the text of the vents and describe the affective landscape of Vent and relate it with an existing emotion lexicon.

\subsection{Emotion Categories and Their Usage}


The main emotion categories in Vent are \emph{Fear}, \emph{Surprise}, \emph{Feelings}, \emph{Sadness}, \emph{Anger}, \emph{Creativity}, \emph{Affection}, \emph{Happiness}, and \emph{Positivity}, some of which match Plutchik's primary emotions \cite{plutchik1991emotions}. 
These emotion categories are always available.
In contrast, there is an additional set of emotion categories that are not always enabled (non-permanent ones) but can become active as in-app purchases, unlocked during a specific season (e.g. Spring, Autumn) or event/festivity (e.g. Women’s Day, Hanukkah).
Other emotion categories can be disabled/deprecated. 
We also identified several (normally disabled) emotions with the same name per category, that can only be differentiated through the provided unique identifiers.

Figure \ref{fig:emo_cats} shows the distribution of vents across the different emotion categories.
For clarity, we show the twenty most used categories from the total of sixty-three categories included in the dataset.
The top ones correspond to the main ones, being~\emph{Sadness} at the top with more than 3M vents, followed by \emph{Feelings} and \emph{Happiness}, with approximately $1.5M$ of vents.
The non-permanent ones (with grey labels) were used less frequently and are grouped at the end of the list.
Despite being less popular, some of the non-permanent ones were extensively used in many vents, e.g., \emph{Springly} or \emph{Valentines17}, which were associated to more than $100K$ vents.

Figure~\ref{fig:emos} shows in more detail the distribution of vents across the different emotions for the top three categories.
We observe a similar pattern regarding enabled/disabled emotions like the one with permanent/non-permanent categories. 
The category \emph{Feelings}, despite being the one containing the largest number of different types of emotions, received many millions of vents less
than \emph{Sadness}.
\begin{figure}[!ht]
  \centering
  \begin{subfigure}[t]{0.49\columnwidth}
  	\centering
\includegraphics[width=\columnwidth]{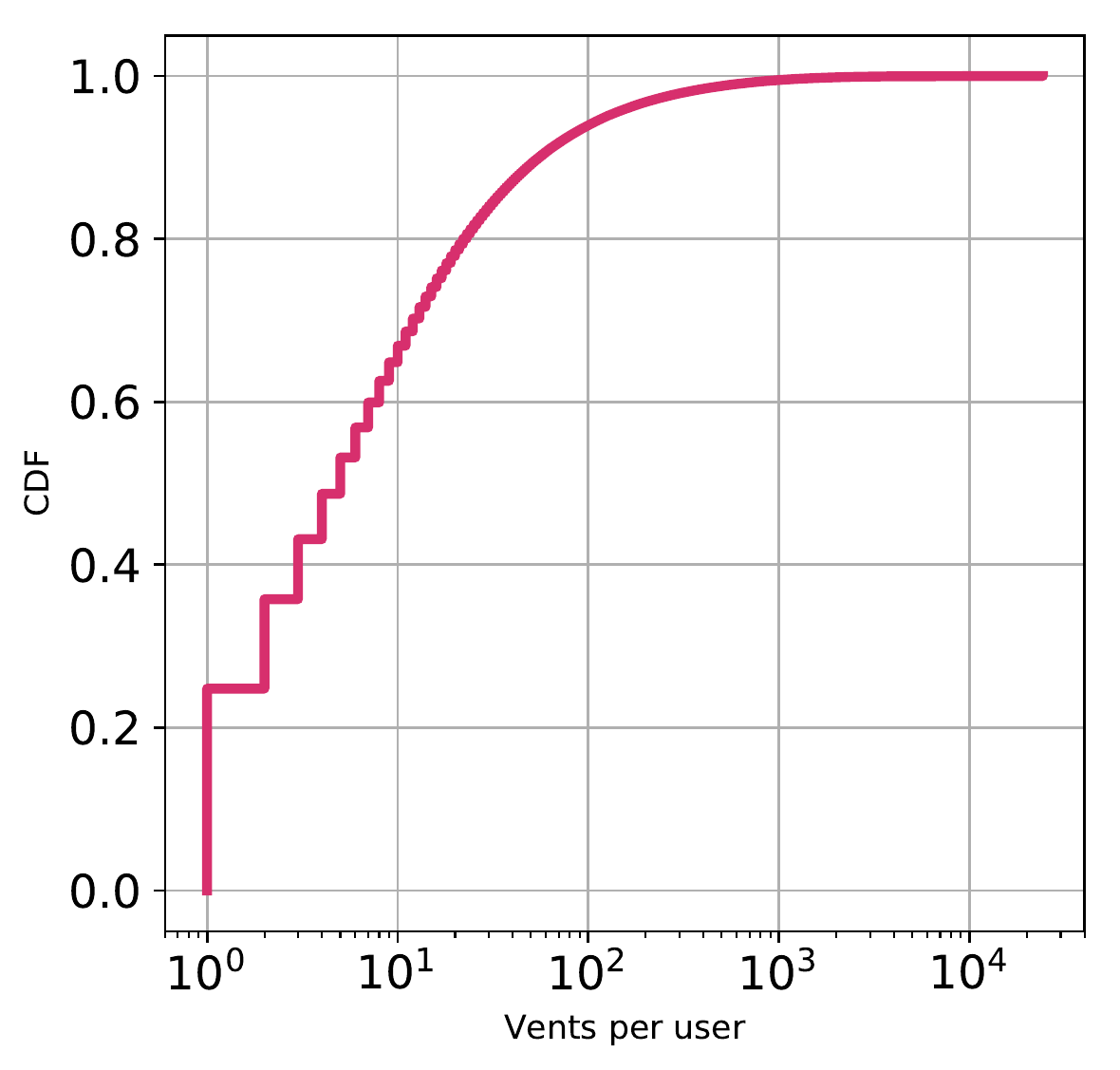}
    \caption{Number of vents per user}
    \label{fig:vents_dist}
  \end{subfigure}
  \begin{subfigure}[t]{0.49\columnwidth}
  	\centering
\includegraphics[width=\columnwidth]{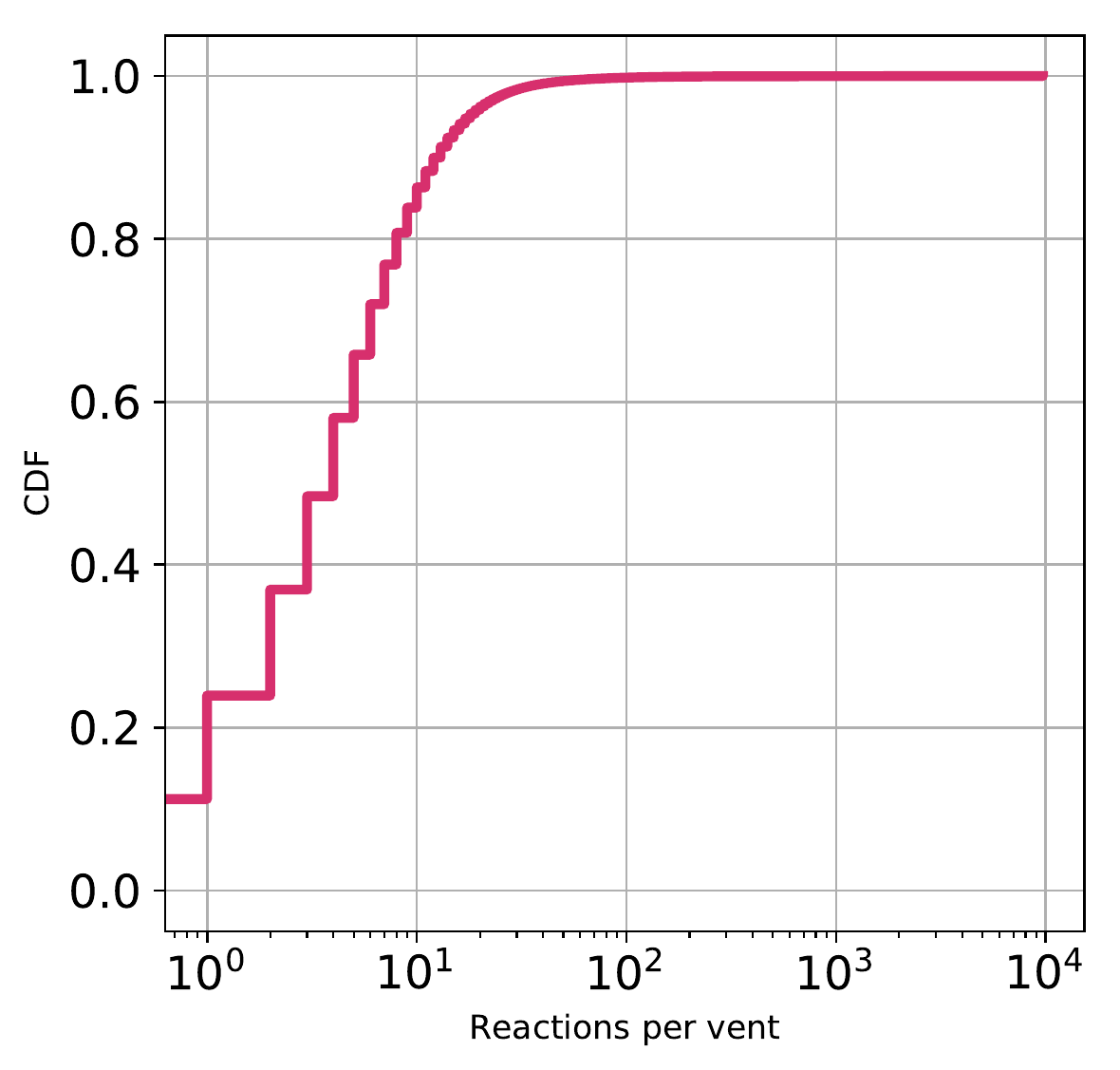}
    \caption{Number of reactions per vent}
    \label{fig:reactions_dist}
  \end{subfigure}  
  \caption{Cumulative distribution functions (CDFs) of the number of vents per user and number of reactions per vent.
Both of them are heavy tailed.}
  \label{fig:ccfs}
\end{figure}

We now examine different aspects of user behavior in the Vent social platform.
Figure~\ref{fig:ccfs} shows the cumulative distribution function (CDF) of the number of vents per user and the number of reactions per vent.
Both of them are governed by heavy-tailed distributions that span four orders of magnitude, indicating that 
a vast majority of $60\%$ of the users posted less than $10$ vents while a small group of few users posted more than $10^4$ vents.
A similar pattern holds for the number of reactions per vent.

\begin{figure}[!ht]
   \begin{subfigure}[t]{0.48\columnwidth}
	\centering
      \includegraphics[width=\columnwidth]{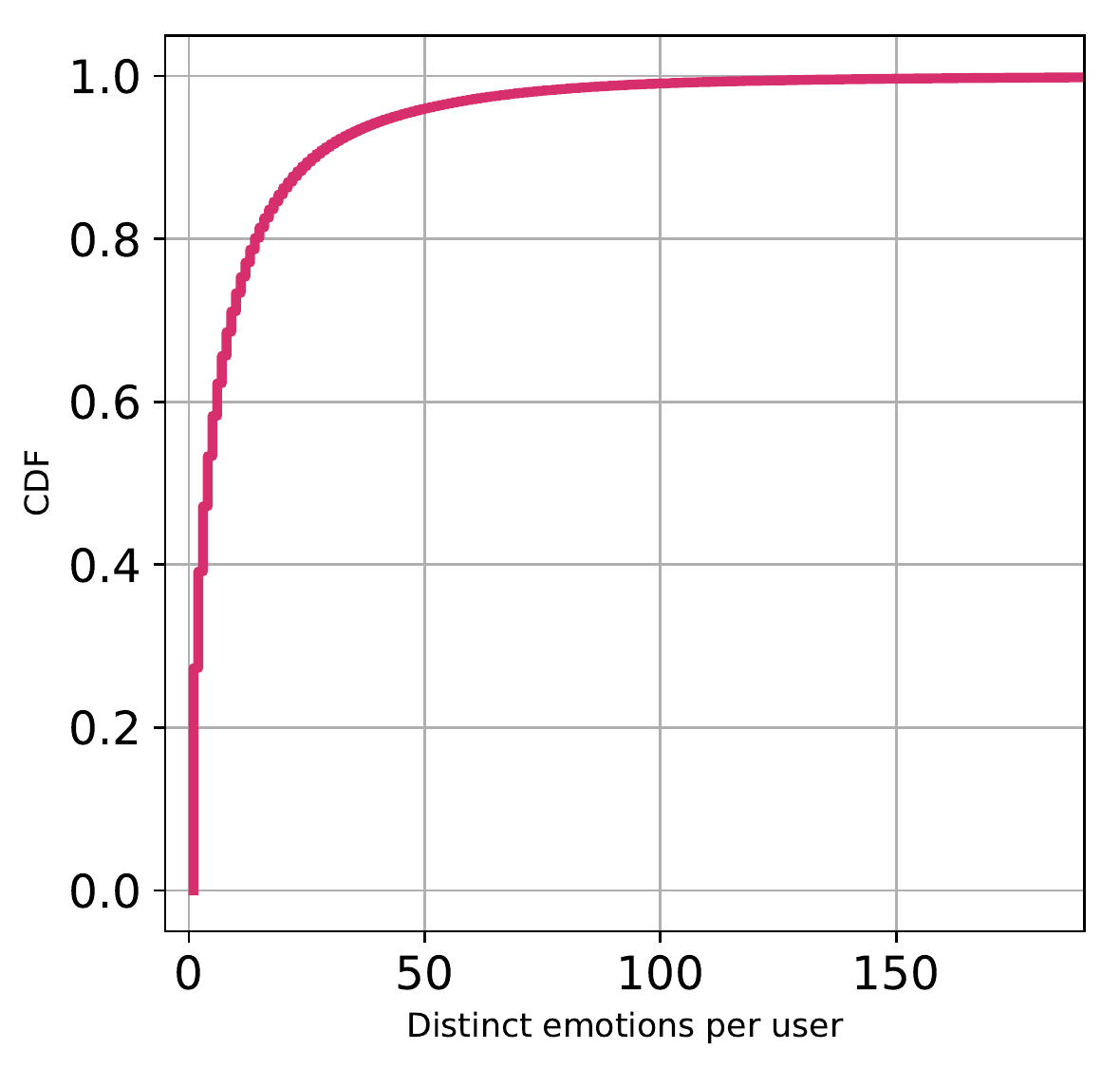}
      \caption{Distinct emotions per user}
      \label{fig:emotions_per_user_cdf}
   \end{subfigure}
       \begin{subfigure}[t]{0.48\columnwidth}
	\includegraphics[width=\columnwidth]{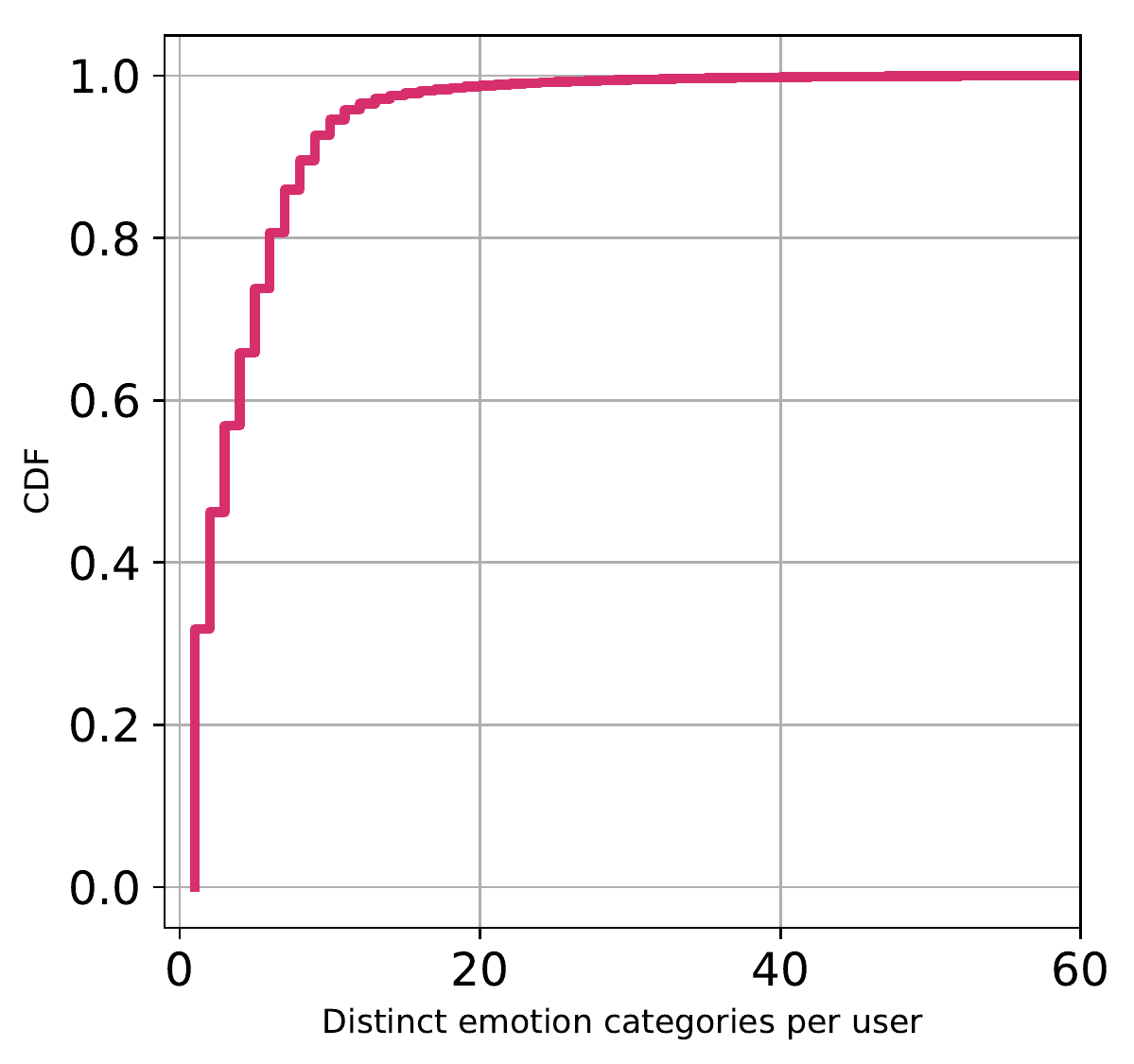}
    \caption{Distinct emotion categories per user}
    \label{fig:emotion_categories_per_user_cdf}
	\end{subfigure}
  \caption{Cumulative distribution functions (CDFs) of the emotions and emotion categories per user.
	Users typically focus on a set of few emotions and emotion categories, but a few users can use a large number of them.}
  \label{fig:CDFs}
\end{figure}

To determine the extent to which venters use the wide range of emotions offered by the Vent platform, we plot the cumulative distribution functions (CDFs) in Figure \ref{fig:CDFs}, for the number of distinct emotions and number of distinct emotion categories associated with the vents of each user.
We observe that the majority of users post vents within a limited set of emotions and emotion categories,
with $50\%$ of users only using up to five different emotions and three emotion categories.
A few users, however, can use up to $60$ emotion categories and more than $200$ emotions.

\subsection{Temporal activity}
\begin{figure}[!h]
  \centering
  \includegraphics[width=\columnwidth]{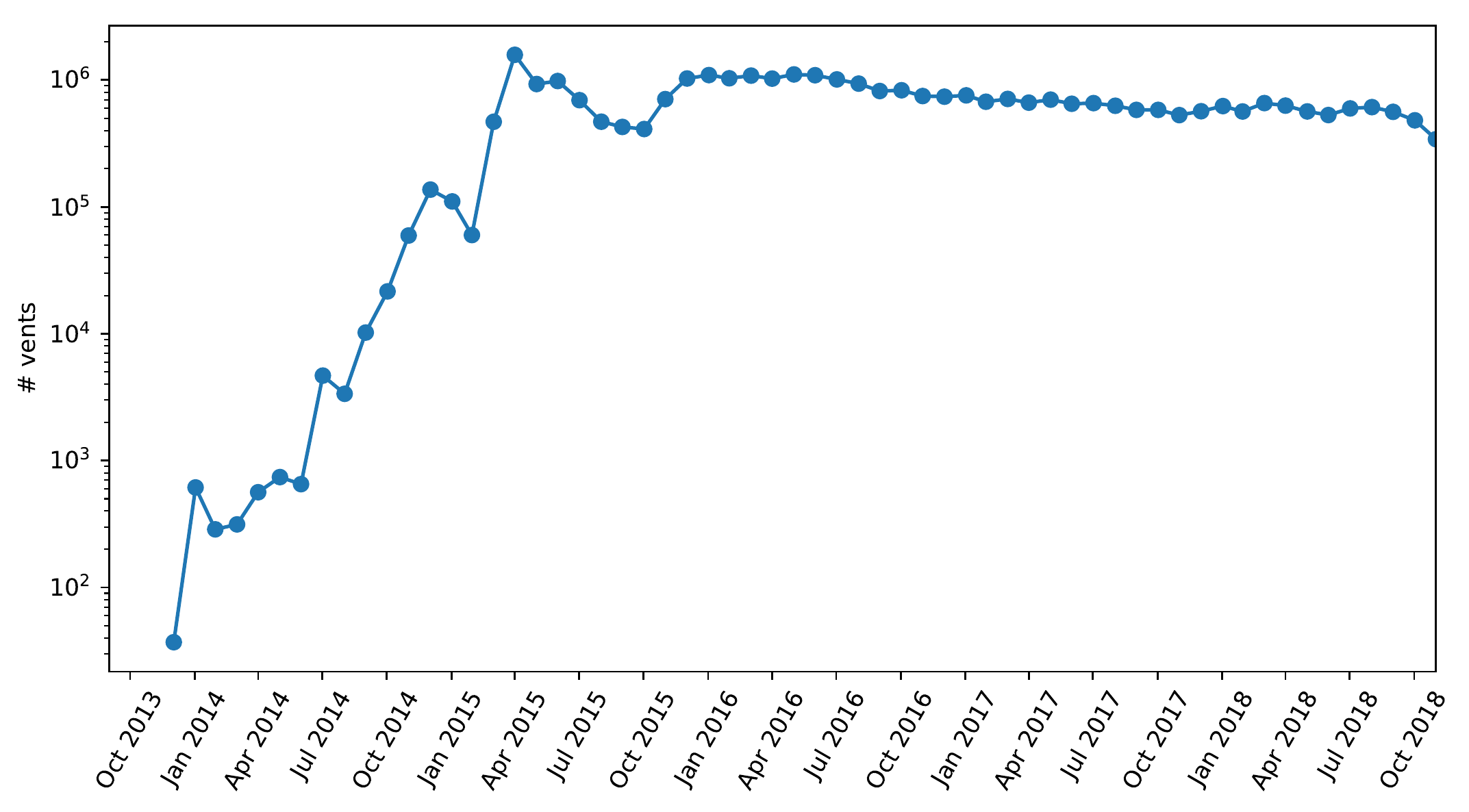}
  \caption{Aggregated monthly activity shows an increase that reached nearly a million of vents .}
  \label{fig:monthly_activity}
\end{figure}

Figure~\ref{fig:monthly_activity} shows the number of vents posted per month, according to the field \texttt{created\_at} in each vent.
The activity shows an increase in the number of vents from December 2013 
(when the Vent app was launched) until a peak of activity was reached around April 2015.
That maximum of activity comprised more than one millions of vents during that month.
Since then, the activity has generally been sustained, but slowly decreasing during the last two years.

\subsection{The social network of Vent}
Vent users can form social links to other users in the platform.
The resulting social network of Vent users has approximately a million of nodes
(users) and contains around 13.5 millions of edges (directional links between them).
Table~\ref{tb:crawled} shows some additional global indicators of the network.

\begin{figure}[t]
  	\centering
   \includegraphics[width=.6\columnwidth]{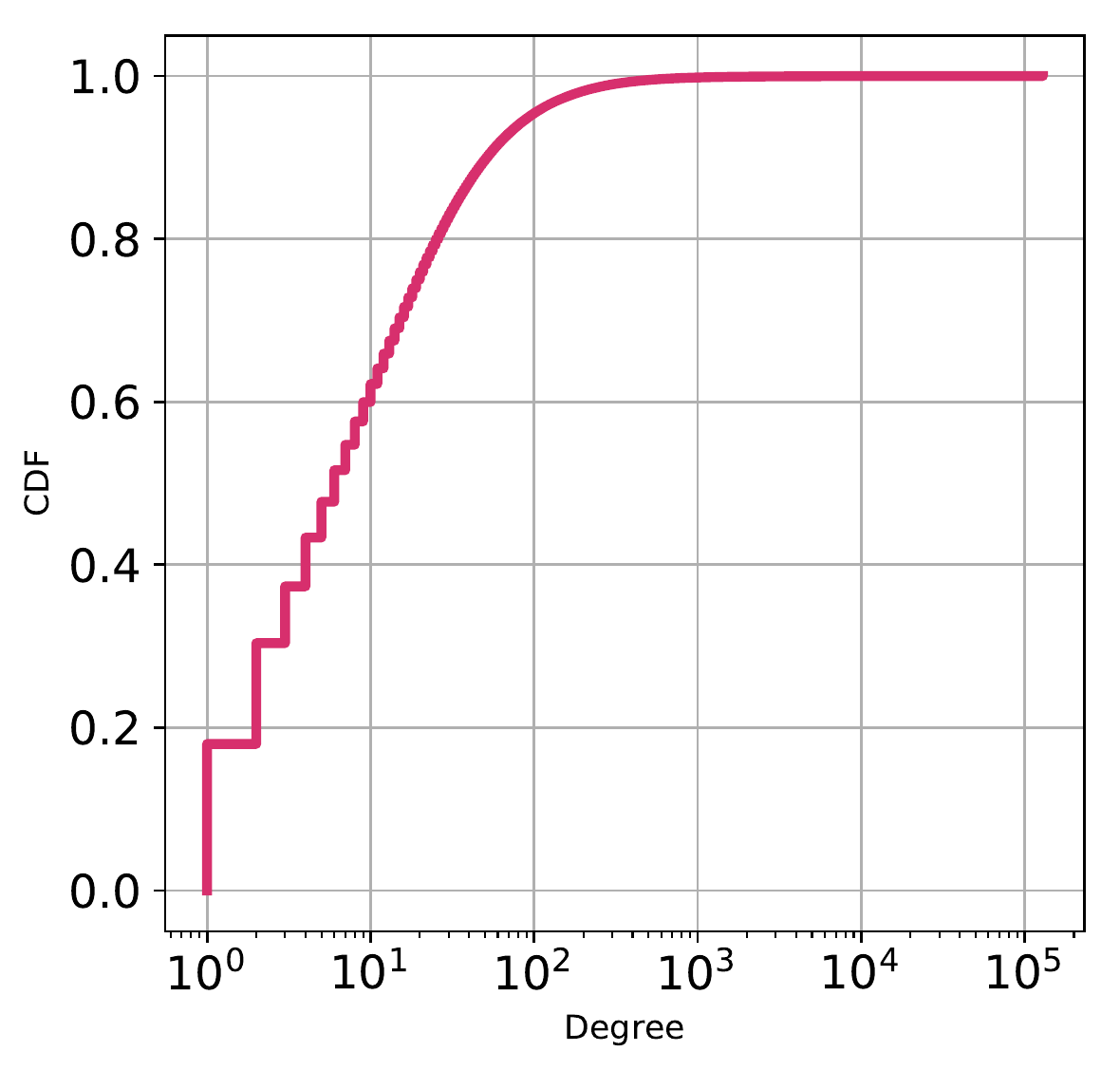}
    \caption{Degree distribution of the social network in Vent.}
    \label{fig:fdegree_dist}
\end{figure}

Figure~\ref{fig:fdegree_dist} shows the degree distribution of the social network. We observe the typical heavy tail behavior, with a vast majority of venters linked to a small number of other users, and a small minority of venters having more than $10K$ links.


\begin{table}[!t]
	\centering
    \caption{Main network indicators  of the Vent social graph: number of nodes $N$, number of edges $E$, average degree $\langle k \rangle$, density $D$, and reciprocity $\rho$.}
     \label{tb:crawled}
      
      \begin{tabular}{rrrcr}
\toprule
$N$ & $E$ & $\langle k \rangle$ & $D$ & $\rho$ \\ \midrule
946,459 & 13,605,522 & 28.7 & 1.51 $\times 10^{-5}$ & 0.53 \\ \bottomrule
\end{tabular}    
\end{table}

\subsection{Text properties of vents}

\begin{figure}[t]
  	\centering
   \includegraphics[width=.6\columnwidth]{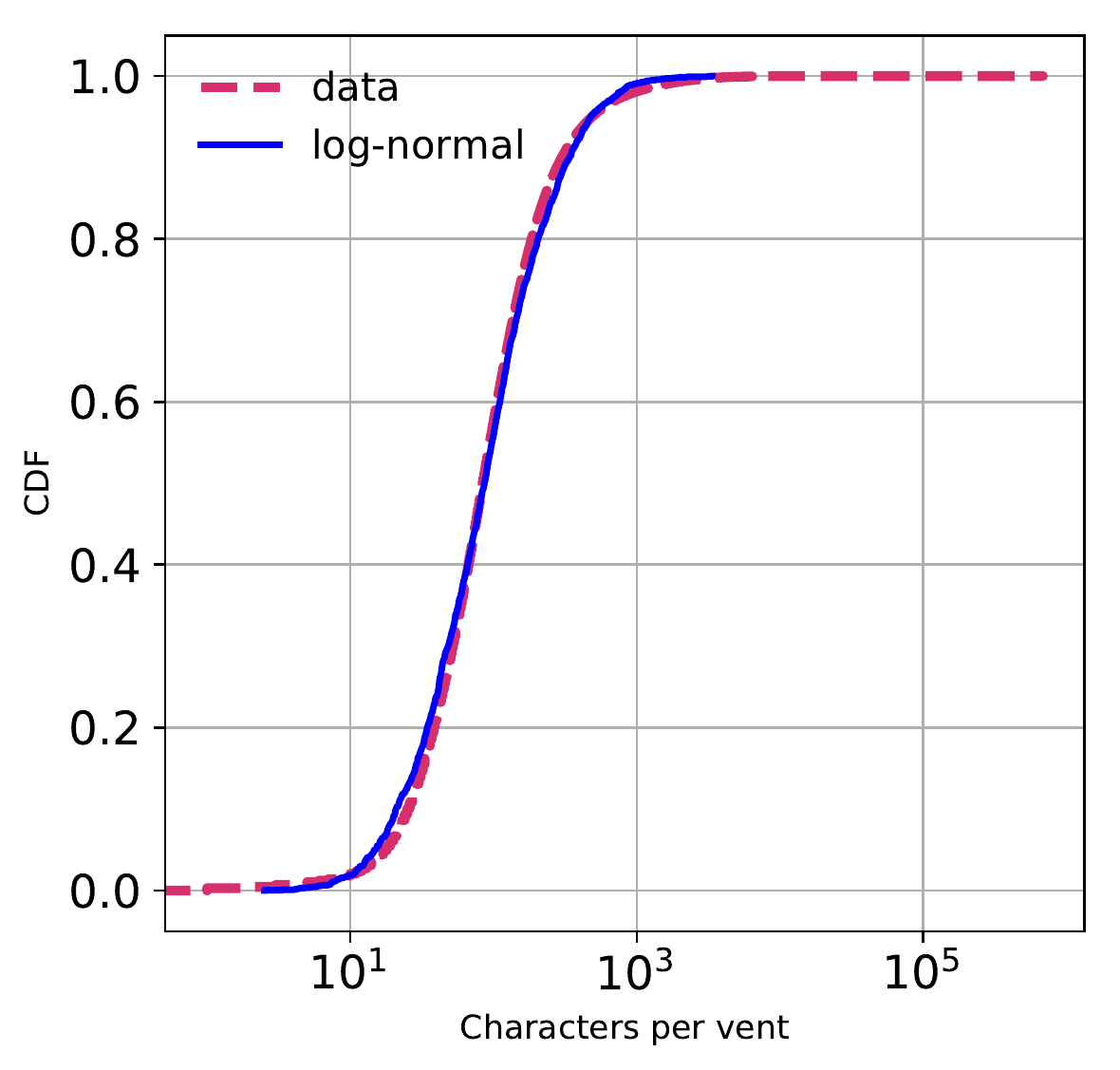}
    \caption{The Vent length (in characters) probability distribution is well approximated by a lognormal ($\mu = 85.57$, $\sigma=1.06$).}
    \label{fig:text_cdf}
\end{figure}

We proceed to examine some of the textual properties of the vents. 
Table~\ref{tab:sample} shows an illustrative sample of vents and their associated emotions.
Figure~\ref{fig:text_cdf} shows the distribution of the vents length, which have a well defined typical length, suggesting 
a log-normal behavior. 
On average, vents are $163.27$ characters long (with a large standard deviation of $388.81$).

In this preliminary analysis, we use an off-the-shelf language identification tool~\cite{lui2012langid} to infer the language of each vent.
As highlighted by the authors, the short size and language novelties of text produced in the context of this type of data 
have a considerable impact on the performance of language recognition. Nonetheless, their model identified 93\% of the vents as English. 

\begin{figure}[!t]
\centering
\begin{subfigure}[t]{\columnwidth}
	\includegraphics[width=0.9\columnwidth]{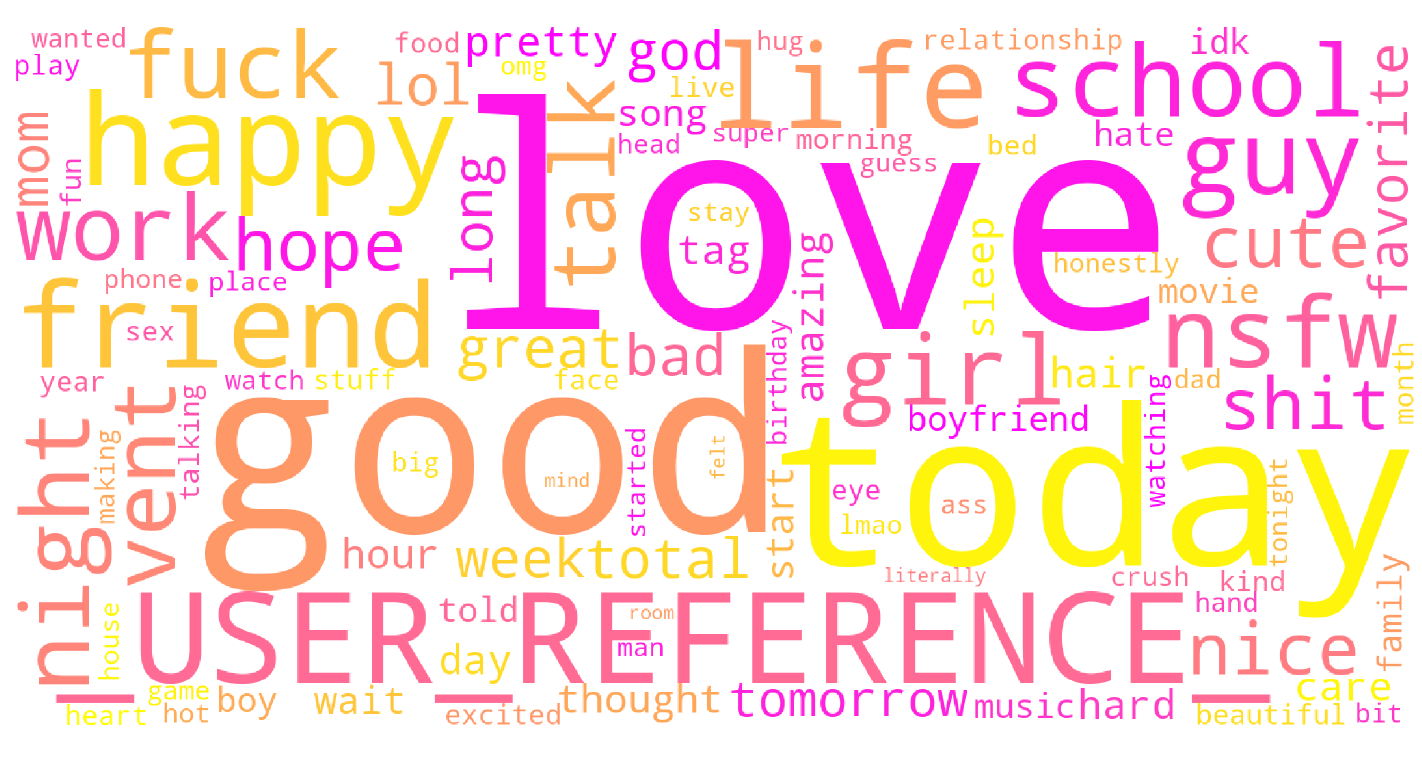}
	\caption{Most frequent words in vents with positive emotions.} 
	\end{subfigure}
	\begin{subfigure}[t]{\columnwidth}
	\includegraphics[width=0.9\columnwidth]{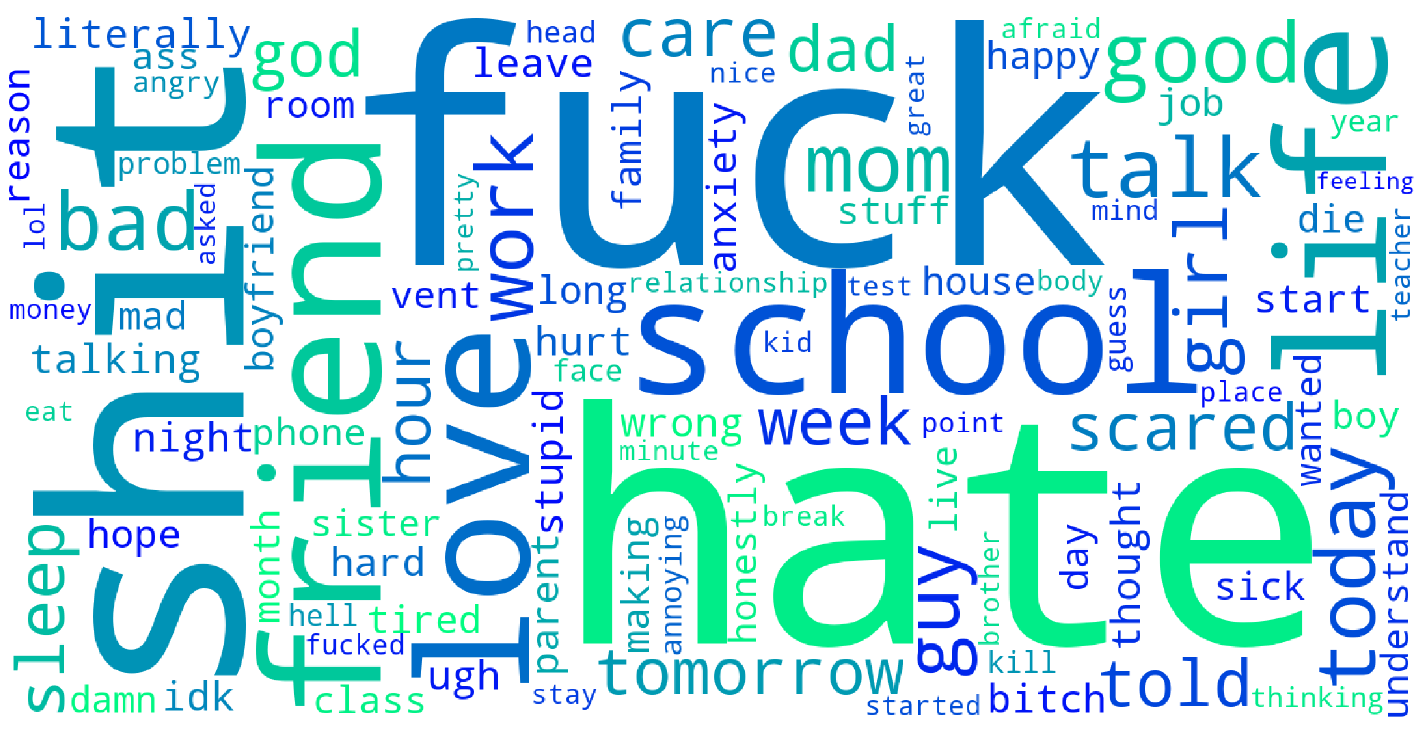}
	\caption{Most frequent words in vents with negative emotions.}
	\end{subfigure}
    \caption{Word clouds built from the text of vents belonging to different (positive and negative) categories.}
    \label{fig:clouds}
\end{figure}

We also generated word clouds from the vents associated with two contrasting groups of emotions: one with emotions belonging to \emph{negative} categories such as \emph{Sadness}, \emph{Anger} and \emph{Fear}, and one with \emph{positive} emotions from \emph{Happiness}, \emph{Affection} and \emph{Positivity} categories.
Figure \ref{fig:clouds} shows these two word clouds generated using \texttt{word\_cloud}\footnote{\url{https://github.com/amueller/word_cloud}}.


Interestingly, one of the predominant words in positive vents appears to be the term \textbf{NSFW}. An explanation from this can be found in the community guidelines of the Vent app\footnote{\url{https://www.vent.co/cg/}}, stating that:
\begin{quote}``\textit{Vent posts that contain sexually explicit content must be flagged with the text `NSFW' in the body of the Vent. NSFW stands for Not Safe For Work and is a common way of describing content that is sexually explicit in nature.}''.\end{quote} 
In total, we identified 508,545 vents in our dataset flagged as explicit. 

\subsection{The network of emotions}
Our dataset can also be used to analyze the relations between emotions.
For example, one can build a network using the following procedure:
let $v(u,e_k)$ the (normalized per user) number of events of user annotated with emotion $e_k$.
We define a pair of emotions $e_i$ and $e_j$ to be related with respect to their common appearance in a particular user's vents, if $v(u,e_i)$ and $v(u,e_j)$ exceed the value of an arbitrary threshold $T_1$ (criterion 1).
Moreover, we can assign a weight to an edge $e_i\sim e_j$ consisting of the number of users satisfying criterion 1 for both emotions, and normalized as the Jaccard similarity between the sets of users satisfying criterion 1 for each emotion.
Additionally, for the sake of visualization, we can control the density of such a graph by filtering out edges with weight below a second threshold $T_2$.
This way the behavior of vent users and, by extension, the affective mental states they express in Vent platform can contribute to the formation of links between different emotions. 

\begin{figure}[!t]
  \centering
  \includegraphics[width=\columnwidth]{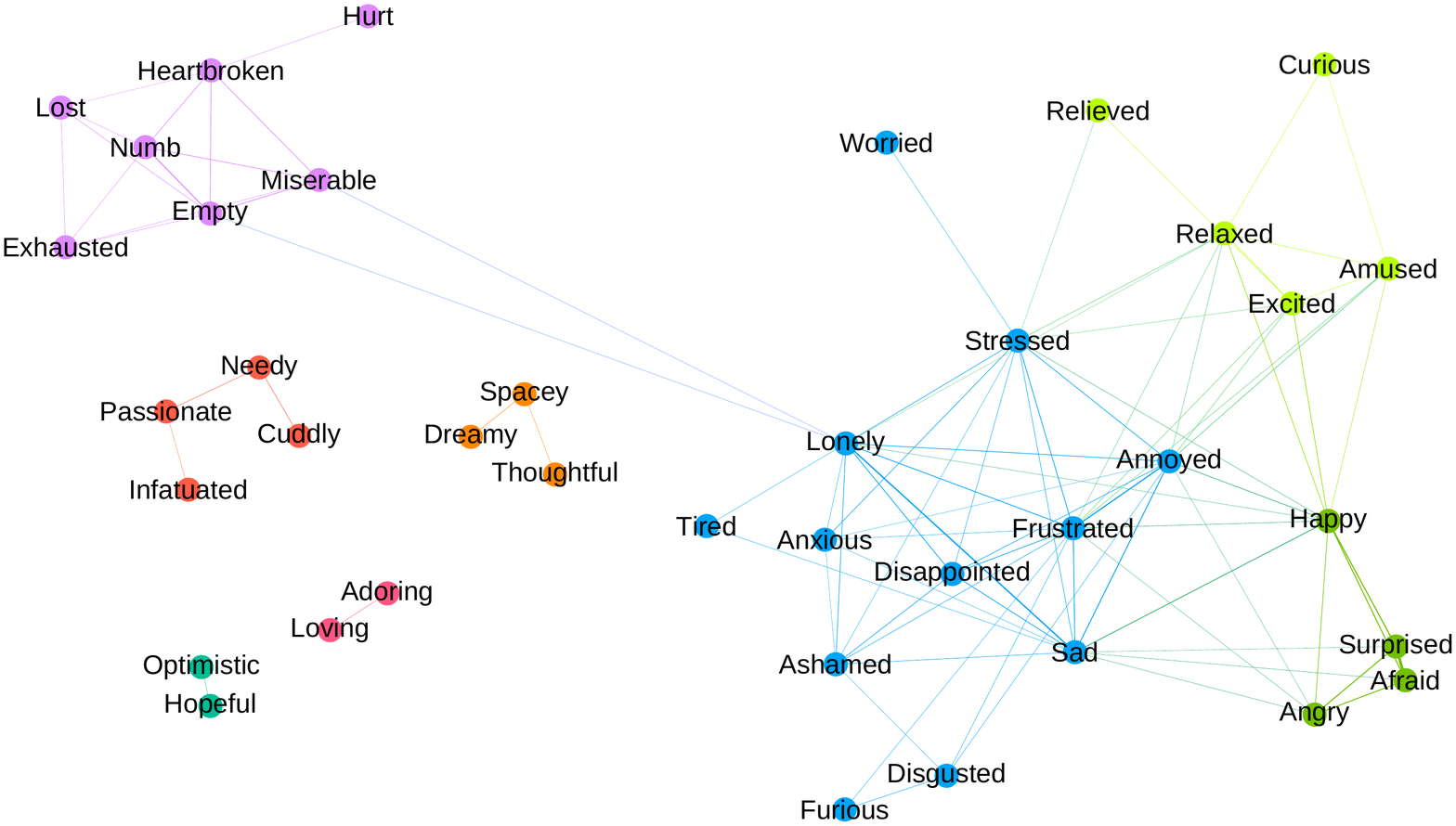} 
  \caption{A example emotion network ($T_1=0.1$ and $T_2=0.05$). See text for details.}
  \label{fig:emo_network}
\end{figure}

Figure~\ref{fig:emo_network} shows the emotion network resulting from applying the steps described above.
Different node colors denote the communities returned by applying the standard Louvain method \cite{blondel2008fast}. 
It is noteworthy that similar emotions are connected, forming distinct clusters, without necessarily belonging to the same emotion category, e.g., \textit{Heartbroken} and \textit{Hurt}.
Moreover, we observe the existence of small connected components, meaning that there exist users that mainly vent about specific moods (such as \textit{Thoughtful}, \textit{Needy}, etc.), outside from the typical spectrum of affect.


From this analysis, one can get some initial insight into the similarity and co-occurrence between different emotions. Our proposed approach can be extended and enriched with the provided temporal information to shed light on complex and largely unexplored affective mechanisms such as the emotion transitions~\cite{thornton2017mental}.

%

\subsection{The ``affective landscape'' of Vent}

We now analyze the text of the vents and explore the ``affective landscape'' of the different emotion categories.
To this end, we use the NRC Emotion Lexicon ``EmoLex'' \cite{mohammad2013nrc}.
The EmoLex dictionary contains 14,182 words crowd-labelled according to the eight Plutchik’s primary emotions \cite{plutchik1991emotions}: ``sadness'', ``joy'', ``disgust'', ``anger'', ``fear'', ``surprise'', ``trust'', and ``anticipation''. Additionally, it includes positive and negative valence categorizations for every included term.


We sampled uniformly at random a set of 3M vents (approximately 10\% of the dataset) and associated each vent to a set of scores per each EmoLex category in the following way: we count the number of words belonging to each EmoLex category, normalize them by the total number of words in the vent, and consider the category-wise mean value. Notably, there are 26\% of the sampled vents for which none of their words was found in EmoLex, suggesting a potential opportunity to extend existing affective lexica such as EmoLex. 
This experiment was repeated multiple times with different random samples of vents and the results that we describe next were consistent across the runs.


\begin{figure}[!t]
\centering
	\includegraphics[width=\columnwidth]{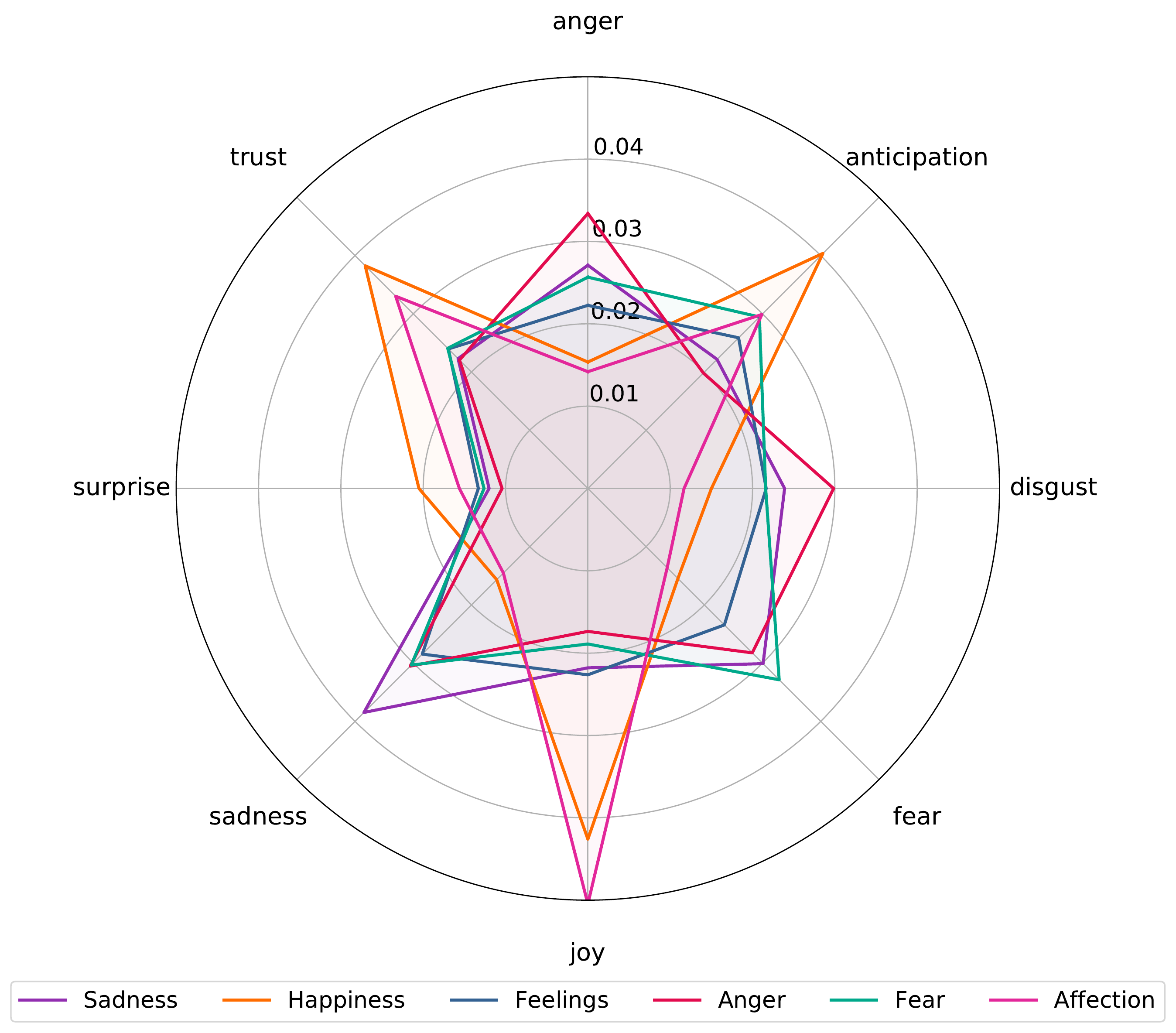}
    \caption{Radar chart visualization of the distribution accross EmoLex categories per each different Vent category (in colours)}
    \label{fig:radar}
\end{figure}

Figure \ref{fig:radar} shows how the top six Vent's emotion categories distribute across each of the EmoLex's emotion categories. 
We observe that vents belonging to the categories of \emph{Happiness} and \emph{Affection} exhibit high levels of ``joy'', ``trust'' and ``anticipation'' while vents of all other categories are dominated by mostly negative EmoLex emotions (``anger'', ``disgust'', ``fear'', sadness'').
Interestingly, EmoLex words from the categories ``anger'', ``fear'' and ``sadness'' are overrepresented in their corresponding Vent categories, as shown in the radar chart visualization. This agreement confirms the alignment and coverage to a large extent of the Vent categories with the ones of the EmoLex annotated corpus.

\begin{figure}[!t]
   \begin{subfigure}[t]{0.495\columnwidth}
	\centering
      \includegraphics[width=\columnwidth]{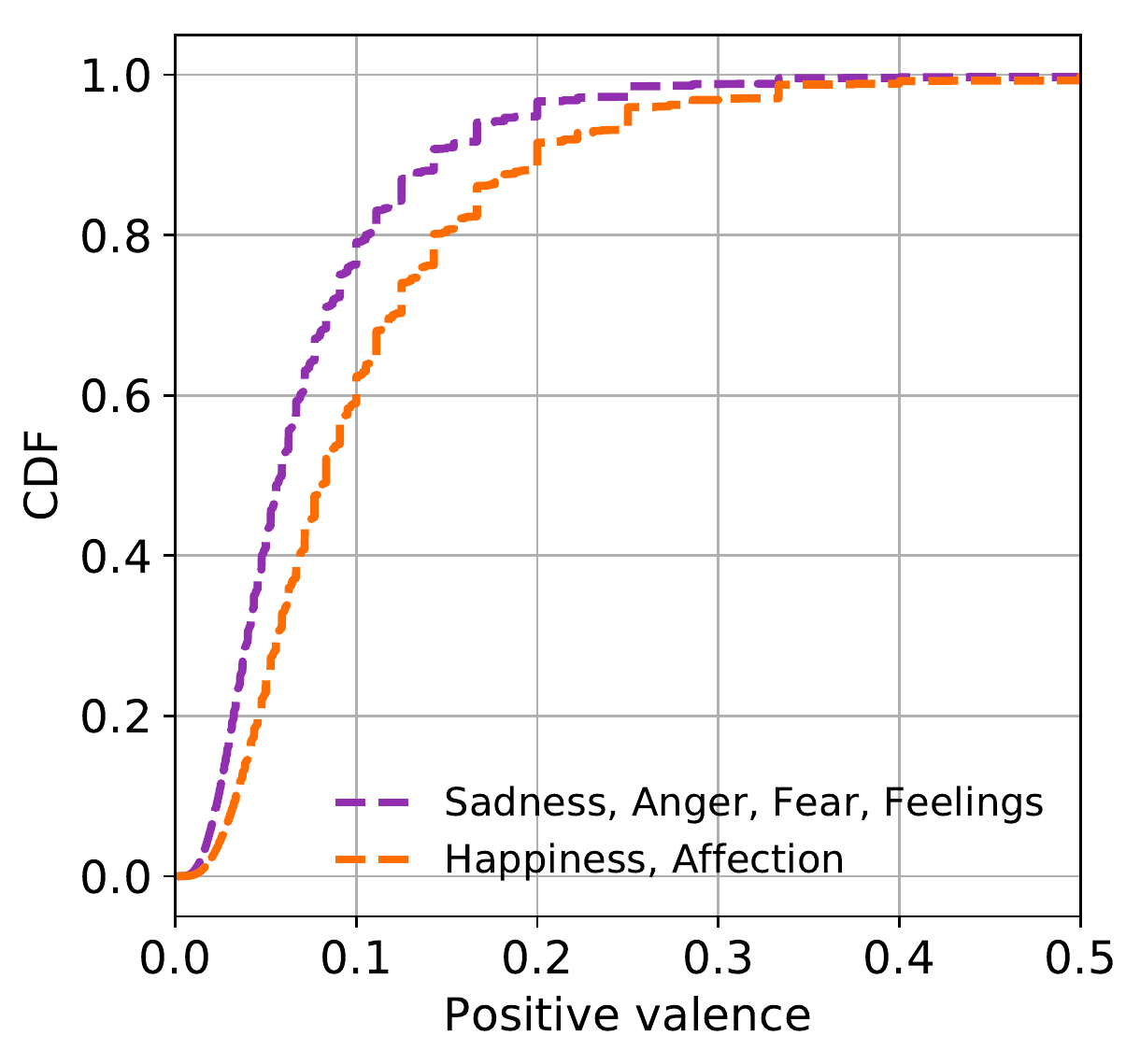}
      \caption{Positive valence}
      \label{fig:pos_valence_cdf}
   \end{subfigure}
       \begin{subfigure}[t]{0.495\columnwidth}
	\includegraphics[width=\columnwidth]{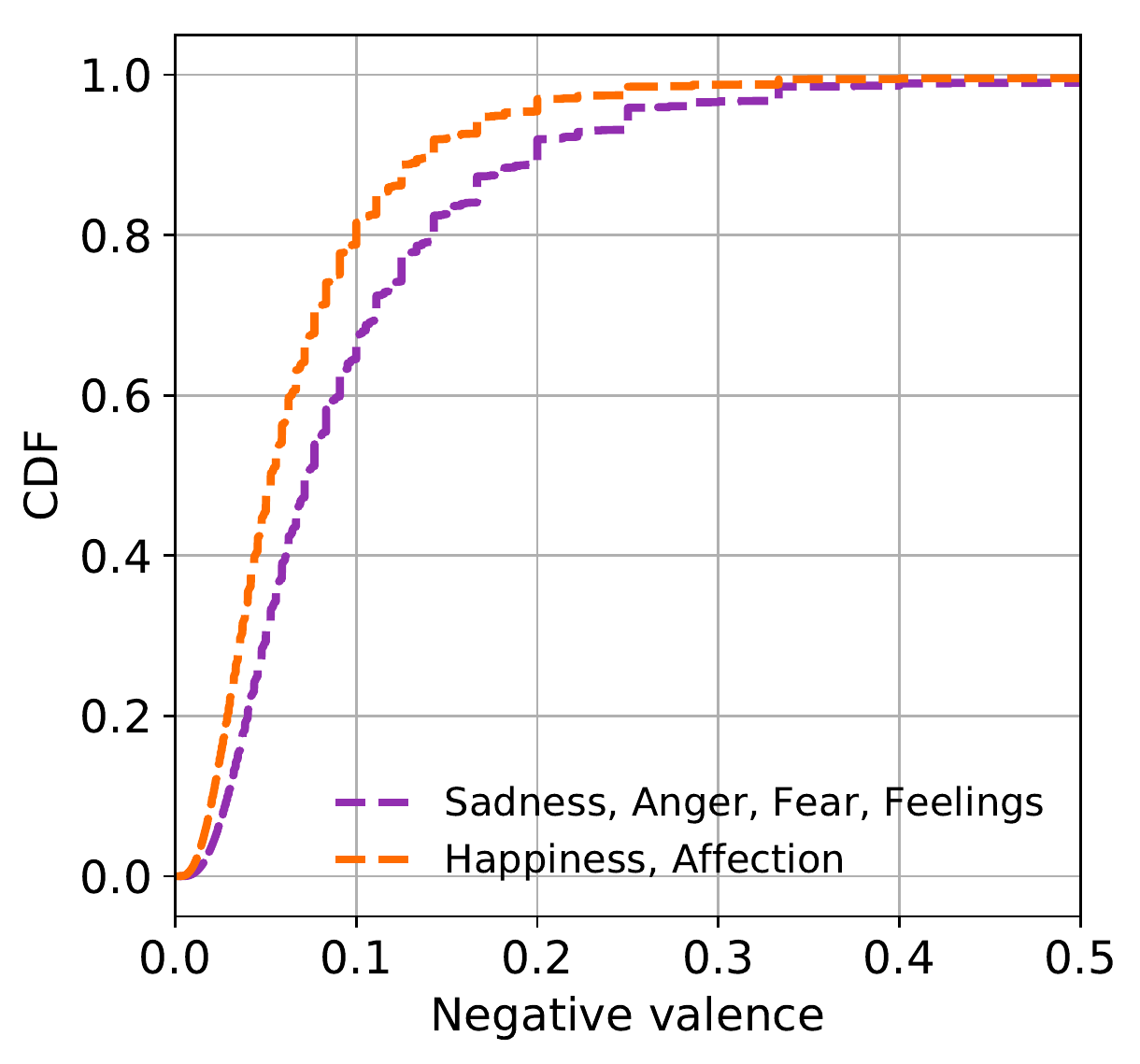}
    \caption{Negative valence}
    \label{fig:neg_valence_cdf}
	\end{subfigure}
  \caption{Cumulative distribution function of the valences grouped by Vent emotion category.
	The positive valences have more probability mass for Vent's categories \emph{Happines} and \emph{Affection}, coinciding with the ``positive'' Vent emotions (CDF is clearly on the right side).
	The opposite holds if we consider negative valences.
}
\end{figure}

Finally, we also analyze the distribution of valence scores (according to EmoLex) for different Vent categories.
We found that the ``positive'' categories of \emph{Happiness} and \emph{Affection} are clearly differentiated from the rest of the categories, which mostly contain negative or neutral emotions.
Figures \ref{fig:pos_valence_cdf} and \ref{fig:neg_valence_cdf} show the CDFs of positive/negative valence scores for the same Vent categories aggregated in two groups and including only nonzero values for readability. 

These results agree with the previous findings, suggesting that there exist substantial differences in emotions expressed in vents across different categories of our dataset.
Also, they highlight certain regularities, despite the large heterogeneity found in the dataset.


\section{Recommendations for Future Work}
In the above section, we have presented a preliminary exploration of our dataset of vents voicing the emotions of approximately 1M users. Given the volume and the content diversity of our dataset, in place of a conclusion, we outline some directions for potential future research.


\subsubsection{Emotion analysis:}
Emotion analysis represents a natural evolution of sentiment analysis. Modeling emotions expressed in text beyond the basic polarity classifications/scales used in sentiment analysis is a challenging task since emotions not only depend on the semantics of a language but are also inherently subjective and ambiguous \cite{sudhof2014sentiment}. Many researchers argue that accounting for affects is crucial in approximating real-world true natural language understanding, especially in areas involving human-computer interactions \cite{park2018plusemo2vec,fung2015robots}.

Recent work has demonstrated that artificial neural networks have great potential in tasks such as emotion recognition \cite{baziotis2018ntua,baziotis2017datastories}. This can be attributed to their ability to learn features directly from data in addition to using hand-crafted features where necessary, thus outperforming conventional approaches which require extensive feature engineering from experts. Such methods, due to their dependence on emotion lexica and hand-crafted features, cannot keep up with rapid language evolution \cite{mudinas2012combining}, especially in social media/micro-blogging context. To this end, we hope the Vent dataset will contribute towards the advancement of emotion analysis in text, by enabling the development and evaluation of novel, neural-network-based models, capable of naturally exploiting its volume and diversity.

\subsubsection{Interplay between network structure and emotional behavior:}
Another line of research exploiting the social graph data of Vent could be the study of social relationships of users with respect to their emotional profiles, and vice-versa.
Previous studies in a variety of social media contexts such as Wikipedia~\cite{iosub2014emotions}, chats~\cite{singla2008yes} and blogs \cite{thelwall2010emotion}, have shown evidence that
users tend to interact and associate with others expressing similar emotions, referred as the phenomenon of ``emotional homophily''. It would be interesting to investigate if relationships among Vent users with respect to the emotions they express adhere to the principle ``birds of a feather flock together''. 

Furthermore, it has been observed that emotions can be diffused through social networks, across the links connecting individuals. Although the phenomenon of emotional contagion is well established in laboratory experiments and real-world social networks \cite{fowler2008dynamic}, in the context of online social networks, it is largely unexplored. Nonetheless, a highly controversial Facebook study \cite{kramer2014experimental} has provided evidence  that emotional states can be transferred to others via emotional contagion, leading people to experience the same emotions without their awareness. Additionally, research in Flickr \cite{yang2016social} has demonstrated that factors such as the social role of individuals are significant with respect to the extent they influence their social connections. We believe the release of the Vent dataset, an example of both network structure (the social network) and diffusion (the vent activity) will stimulate research leading to better understanding the underlying mechanism of emotional contagion in online social networks.

\section{Acknowledgements}
This work was supported by the European Commission under the Horizon 2020 Programme (H2020), as part of the~\href{http://practicies.org/}{Practicies} project (Grant Agreement no. 740072). We also thank NVIDIA Corporation for their GPU donation supporting our research.

\bibliographystyle{aaai}
\bibliography{references}
\balance
\end{document}